\documentclass[11pt,aps,showpacs,superscriptaddress]{revtex4-1}
\usepackage{graphicx}
\usepackage{amsfonts,amssymb}
\usepackage{enumerate}
\usepackage[sort&compress]{natbib}
\usepackage{hyperref}
\usepackage[latin1]{inputenc}

\usepackage{amsfonts,amsmath,amssymb,graphicx,color,amsthm}
\usepackage{tgtermes}
\usepackage{pdfpages}

\makeatletter
\AtBeginDocument{\let\LS@rot\@undefined}
\makeatother

\def\EE{\mathbb{E}}
\def\PP{\mathbb{P}}
\def\RR{\mathbb{R}}

\newcommand{\eve}[1]{#1}
\newcommand{\tg}[1]{#1}
\newcommand{\gd}[1]{#1}

\DeclareRobustCommand{\argmin}{\operatorname*{argmin}}

\begin{document}

\title{Rogue Waves and Large Deviations in Deep Sea}

\author{Giovanni \surname{Dematteis}}
\affiliation{Courant Institute,
  New York University, 251 Mercer Street, New York, NY 10012, USA}
\affiliation{Dipartimento di Scienze Matematiche, Politecnico di
  Torino, Corso Duca degli Abruzzi 24, I-10129 Torino, Italy}

\author{Tobias \surname{Grafke}}
\affiliation{Mathematics Institute, University of Warwick, Coventry CV4 7AL, United Kingdom} 

\author{Eric \surname{Vanden-Eijnden}} 
\affiliation{Courant Institute, New York University, 251 Mercer
  Street, New York, NY 10012, USA}

\date{\today}

\begin{abstract}
  The appearance of rogue waves in deep sea is investigated using the
  modified nonlinear Schr\"odinger (MNLS) equation in one
    spatial-dimension with random initial conditions that are assumed
  to be normally distributed, with a spectrum approximating
    realistic conditions of a uni-directional sea state. It is shown
    that one can use the incomplete information contained in this
    spectrum as prior and supplement this information with the MNLS
    dynamics to reliably estimate the probability distribution of the
    sea surface elevation far in the tail at later times. Our results
    indicate that rogue waves occur when the system hits unlikely
    pockets of wave configurations that trigger large disturbances of
    the surface height. The rogue wave precursors in these pockets
  are wave patterns of regular height but with a very specific shape
  that is identified explicitly, thereby allowing for early
  detection. The method proposed here combines Monte Carlo
    sampling with tools from large deviations theory that reduce the
  calculation of the most likely rogue wave precursors to an
  optimization problem that can be solved efficiently. This approach
  is transferable to other problems in which the system's governing
  equations contain random initial conditions and/or parameters.
\end{abstract}

\maketitle
Rogue waves, long considered a figment of sailor's
imagination, are now recognized to be a real, and serious, threat for
boats and naval structures~\cite{muller2005rogue,
  white1998chance}. Oceanographers define them as deep water waves
whose crest-to-trough height $H$ exceeds twice the significant wave
height $H_s$, which itself is four times the standard deviation of the
ocean surface elevation.  Rogue waves appear suddenly and
unpredictably, and can lead to water walls with vertical size on the
order of $20$--$30$~m~\cite{haver2004possible,nikolkina2011rogue},
with enormous destructive power. Although rare, they tend to occur
more frequently than predicted by linear Gaussian
theory~\cite{onorato-residori-bortolozzo-etal:2013,nazarenko:2016}. While
the mechanisms underlying their appearance remain under
debate~\cite{akhmediev2009extreme,akhmediev2010editorial,onorato2016origin},
one plausible scenario has emerged over the years: it involves the
phenomenon of modulational
instability~\cite{benjamin-feir:1967,zakharov:1968}, a nonlinear
amplification mechanism by which many weakly interacting waves of
regular size can create a much larger one. Such an instability arises
in the context of the focusing nonlinear Schr\"odinger (NLS)
equation~\cite{zakharov:1968, kuznetsov:1977, peregrine:1983,
  akhmediev1987exact, osborne2000nonlinear, zakharov-ostrovsky:2009,
  onorato:2009} or its higher order variants~\cite{dysthe:1979,
  stiassnie:1984, trulsen1996modified, craig2010hamiltonian,
  gramstad2011hamiltonian}, which are known to be good models for the
evolution of a unidirectional, narrow-banded surface wave field in a
deep sea. Support for the description of rogue waves through such
envelope equations recently came from experiments in water
tanks~\cite{onorato2004observation, chabchoub2011rogue,
  chabchoub2012super, goullet2011numerical}, where Dysthe's MNLS
equation in one spatial dimension~\cite{dysthe:1979, stiassnie:1984}
was shown to accurately describe the mechanism creating coherent
structures which soak up energy from its surroundings. While these
experiments and other theoretical works~\cite{lo1985numerical,
  cousins-sapsis:2015A} give grounds for the use of MNLS to describe
rogue waves, they have not addressed the question of their likelihood
of appearance. Some progress in this direction has been recently made
in~\cite{cousins2016reduced}, where a reduced model based on MNLS was
used to estimate the probability of a given amplitude within a certain
time, and thereby compute the tail of the surface height
distribution. These calculations were done using an ansatz for the
solutions of MNLS, effectively making the problem two-dimensional. The
purpose of this paper is to remove this approximation, and study the
problem in its full generality. Specifically, we consider the MNLS
with random initial data drawn from a Gaussian
distribution~\cite{nazarenko2011wave}. The spectrum of this field is
chosen \eve{to have a width comparable to that of the JONSWAP
  spectrum~\cite{hasselmann:1973,onorato:2001} obtained from
  observations in the North Sea.}  We calculate the probability of
occurrence of a large amplitude solution of MNLS out of these random
initial data, and thereby also estimate the tail of the surface height
distribution. These calculations are performed within the framework of
large deviations theory (LDT), which predicts the most likely way by
which large disturbances arise and therefore also explains the
mechanism of rogue wave creation. \eve{Our results are validated by
  comparison with brute-force Monte-Carlo simulations, which indicate
  that rogue waves in MNLS are indeed within the realm of LDT. Our
  approach therefore gives an efficient way to assess the probability
  of large waves and their mechanism of creation.}

\section{Problem setup} 

Our starting point will be the MNLS equation for the evolution of the
complex envelope $u(t,x)$ of the sea surface in deep
water~\cite{dysthe:1979}, in terms of which the surface elevation
reads $\eta(t,x) = \Re\big(u(t,x)e^{i(k_0x -\omega_0 t)}\big)$ (here
$k_0$ denotes the carrier wave number, $\omega_0= \sqrt{gk_0}$, and
$g$ is the gravitational acceleration). Measuring~$u$ and~$x$ in units
of~$k^{-1}_0$ and~$t$ in $\omega_0^{-1}$ we can write MNLS in
non-dimensional form as
\begin{equation}
  \label{eq:MNLSnondim}
  \begin{aligned}
    &\partial_{t} u + \tfrac{1}{2}\partial_{x} u +
    \tfrac{i}{8} \partial^2_{x}u -\tfrac{1}{16} \partial^3_{x}u
    +  \tfrac{i}{2}|u|^2u \\
    &+ \tfrac{3}{2}|u|^2\partial_x u + \tfrac{1}{4}u^2 \partial_x \bar
    u - \tfrac{i}{2} \left|\partial_x\right| |u|^2 = 0, \quad x\in [0,L],
\end{aligned}
\end{equation}
where the bar denotes complex conjugation.  We will consider
Eq.~\ref{eq:MNLSnondim} with random initial condition
\mbox{$u_0(x)\equiv u(0,x)$}, constructed via their Fourier
representation,
\begin{equation}
  \label{eq:1}
  u_0(x) = \sum_{n\in \mathbb{Z}}  e^{ik_n x}
  (2\hat{C}_n)^{1/2}\theta_n, \quad \hat{C}_n =  A
    e^{-k_n^2/(2\Delta^2)},
\end{equation}
where $k_n=2\pi n/L$, $\theta_n$ are complex Gaussian variables with
mean zero and covariance
\mbox{$\EE\theta_n\bar\theta_m=\delta_{m,n}$},
$\EE \theta_n \theta_m = \EE\bar \theta_n \bar \theta_m = 0$. This
guarantees that $u_0(x)$ is a Gaussian field with mean zero and
$\EE (u_0(x)\bar u_0(x')) = 2\sum_{n\in \mathbb{Z}} e^{ik_n (x-x')}
\hat C_n$.To make contact with the observational data, the amplitude
$A$ and the width $\Delta$ in Eq.~\ref{eq:1} are picked so that
$\hat C_n$ has the same height and area as the JONSWAP
spectrum~\cite{hasselmann:1973,onorato:2001} \eve{-- see
  \textit{Supporting Information} for details}.

Because the initial data for Eq.~\ref{eq:MNLSnondim} are random, so is
the solution at time $t>0$, and our aim is to compute
\begin{equation}
\label{eq:expectations}
 P_T(z) \equiv \PP \,\big(F(u(T))\ge z\big), 
\end{equation}
where $\PP$ denotes probability over the initial data and $F$ is a
scalar functional depending on $u$ at time $T>0$. Even though our
method is applicable to more general observables, here we will focus
on
\begin{equation}
\label{eq:observable}
 F(u(T)) = \max_{x\in[0,L]} |u(T,x)|.
\end{equation}
%

\begin{figure}
  \centering
  \includegraphics[width=0.6\columnwidth]{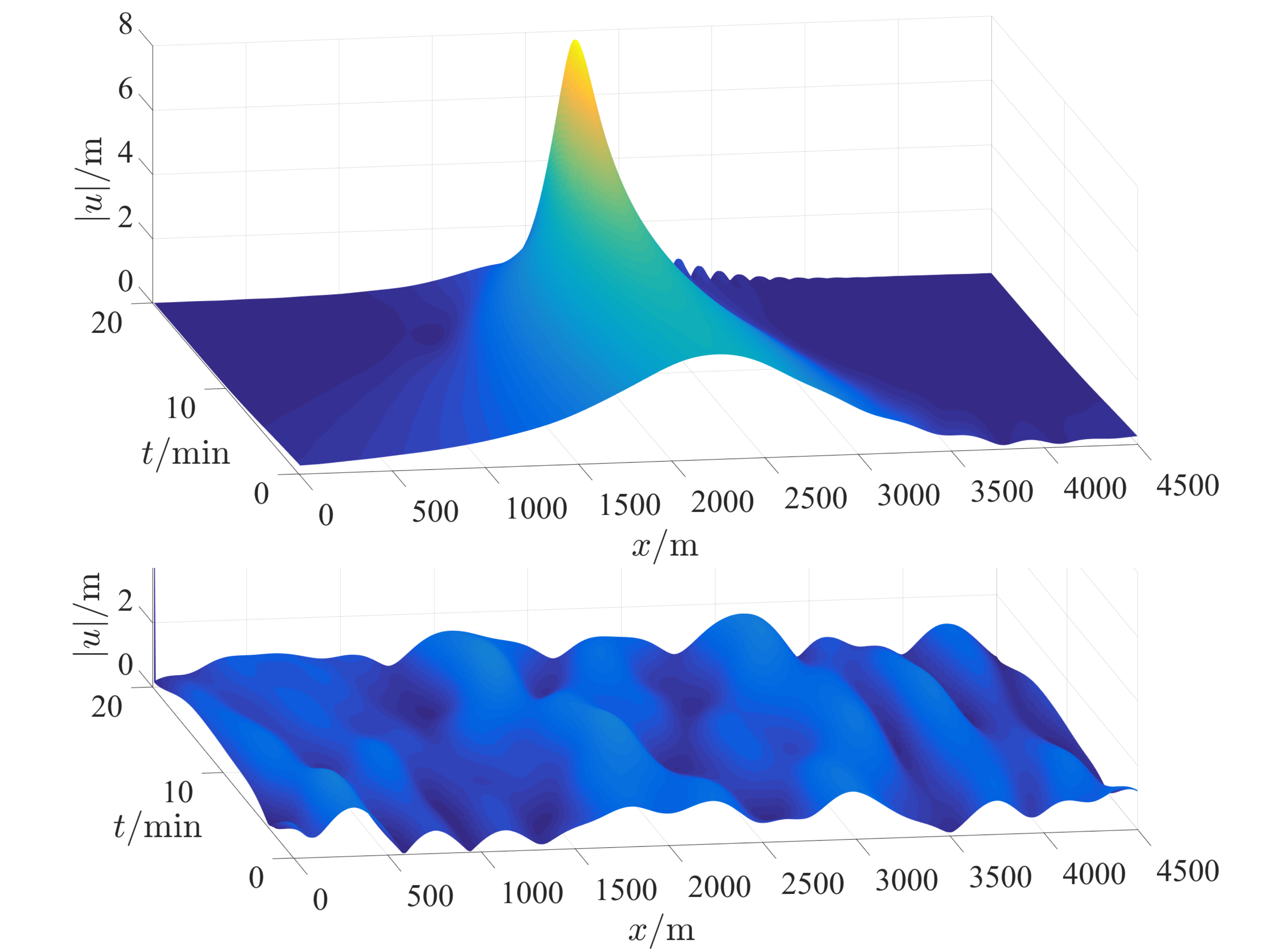}
  \caption{\label{fig:1} \textit{Top:} Time evolution of $|u(t,x)|$
    from an initial condition optimized for $\max_x |u(T,x)|\ge 8$~m
    at $T=20$~min. \textit{Bottom:} Same for a typical Gaussian random
    initial condition. }
\end{figure}

\begin{figure}
  \begin{center}
    \includegraphics[width=0.6\columnwidth]{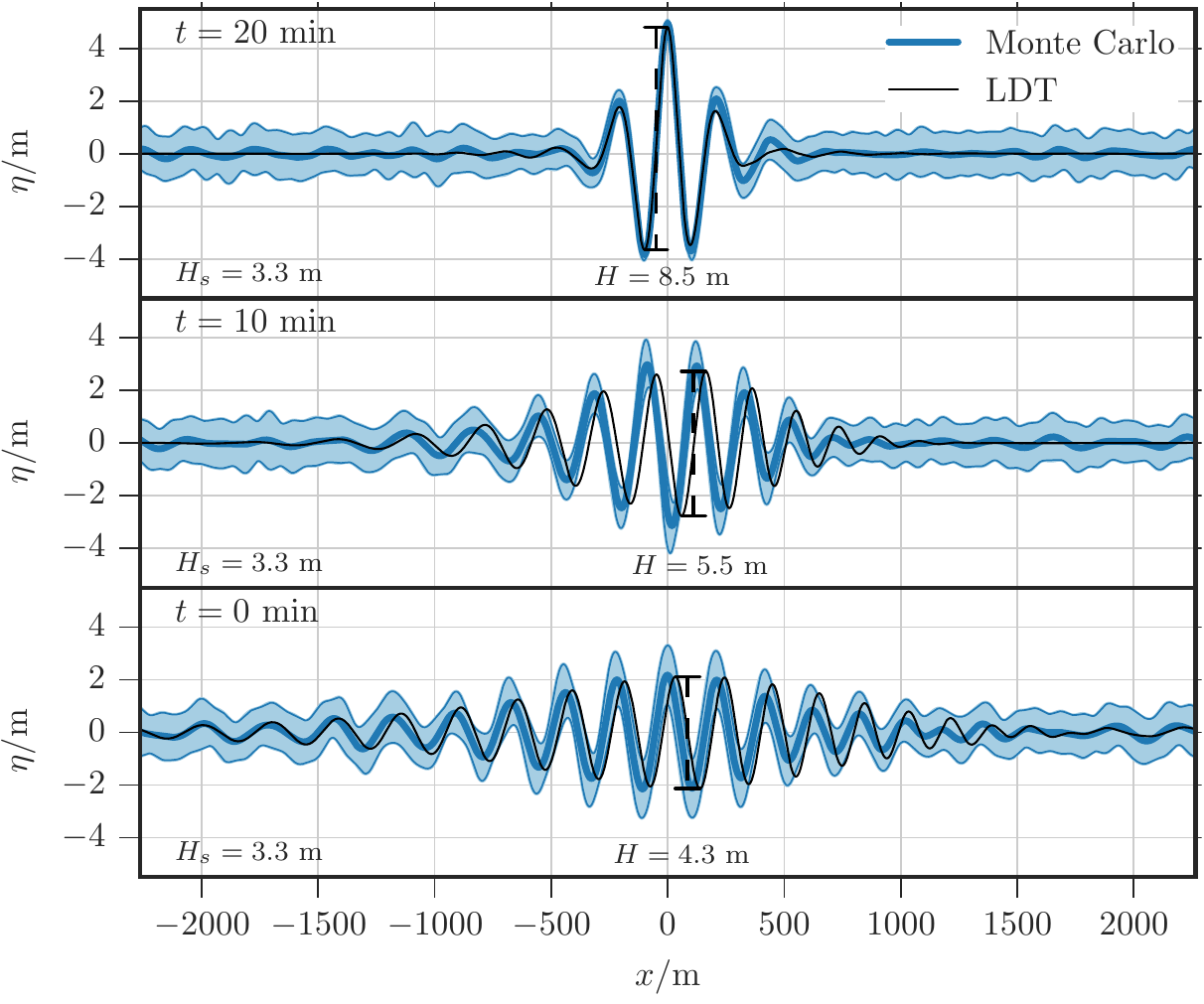}
  \end{center}
  \caption{\label{fig:2} Comparison between the average realization
    reaching $\max_x |u(T,x)|\ge 4.8$~m at $T=20$ min (dark blue) and
    one standard deviation around this mean (light blue), with the
    solution reaching the same amplitude starting from the maximum
    likelihood initial condition (black) for $t=0,10,20$ min.}
\end{figure}

\begin{figure*}
  \begin{center}\includegraphics[width=\textwidth]{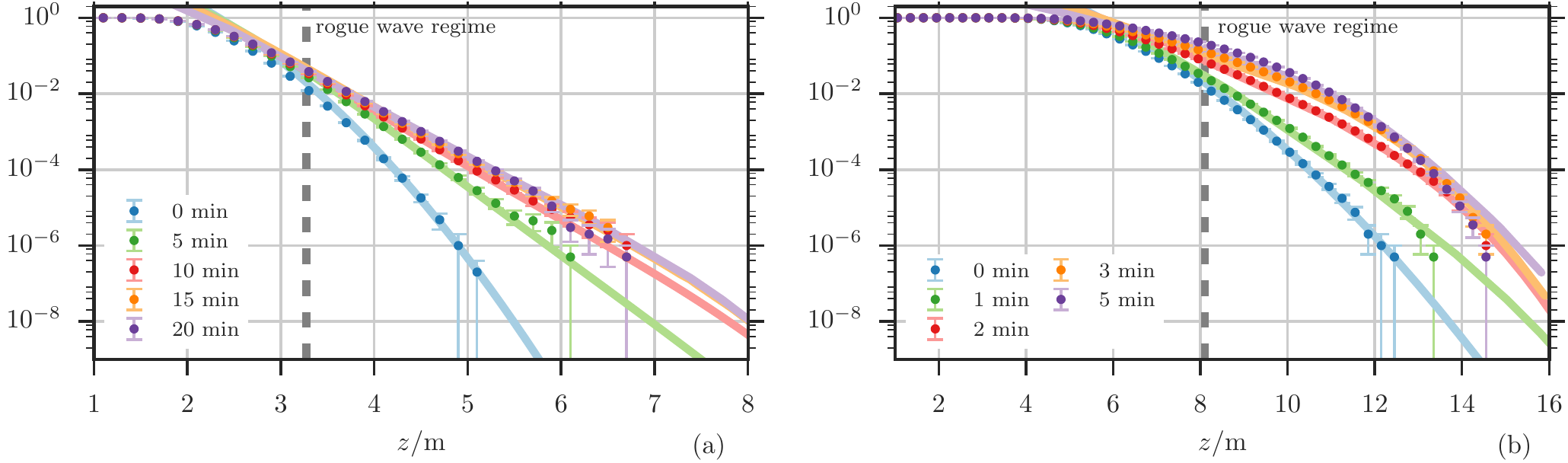}
  \end{center}
  \caption{\label{fig:3} \eve{Probabilities $P_T(z)$ of
      $\max_x |u(T,x)|$ for different times $T$ for Set~1 (a) and
      Set~2 (b). }The probabilities estimated by Monte-Carlo sampling
    with $10^6$ realizations (dots) are compared with those predicted
    by LDT (lines). These probabilities agree over about 5 orders of
    magnitude in probability, though LDT allows for the calculation of
    the tail where Monte-Carlo becomes prohibitively costly. \eve{The
      error bars in the Monte-Carlo results represent the statistical
      error of 2 standard deviations ($95\%$ confidence interval) for
      the Bernoulli distribution with parameter $P_T(z)$}.}
\end{figure*}

\section{Large deviations theory approach}

A brute force approach to calculate Eq.~\ref{eq:expectations} is
Monte-Carlo sampling: Generate \eve{random initial conditions $u_0(x)$
  by picking random $\theta_n$'s in Eq.~\ref{eq:1}, evolve each of
  these $u_0(x)$ deterministically via Eq.~\ref{eq:MNLSnondim} up to
  time $t=T$ to get $u(T,x)$, and count the proportion that fulfill
  $F(u(T))\ge z$. While this method is simple, and will be used below
  as benchmark, it looses efficiency when $z$ is large, which is
  precisely the regime of interest for the tails of the distribution
  of $F(u(T))$.} In that regime, a more efficient approach is to rely
on results from LDT which assert that Eq.~\ref{eq:expectations} can be
estimated by identifying the most likely initial condition that is
consistent with $F(u(T))\ge z$. To see how this result comes about,
recall that the probability density of $u_0$ is formally proportional
to $\exp(-\frac12 \|u_0\|^2_C)$, where $\|u_0\|^2_C$ is given by
\begin{equation}
  \label{eq:3}
  \| u_0\|^2_C = \sum_{n\in \mathbb{Z}} \frac{|\hat a_n|^2}{\hat C_n},
  \quad \hat a_n = \frac1{L} \int_0^{L} e^{-ik_n x} u_0(x) dx\,.
\end{equation}
To calculate Eq.~\ref{eq:expectations} we should integrate this
density over the set $\Omega(z)=\{u_0: F(u(T,u_0))\ge z\}$, which is
hard to do in practice. Instead we can estimate the integral by
Laplace's method. As shown in \textit{Material and Methods}, this is
justified for large~$z$, when the probability of the set $\Omega(z)$
is dominated by a single $u_0(x)$ that contributes most to the
integral and can be identified via the constrained
minimization problem
\begin{equation}
  \label{eq:6}
  \tfrac12\min_{u_0\in \Omega(z)}\, \| u_0\|^2_C\equiv I_T(z)\,,
\end{equation}
which then yields the following LDT estimate for Eq.~\ref{eq:expectations}
\begin{equation}
  \label{eq:4}
  P_T(z) \asymp \exp\left(-I_T (z) \right)\,.
\end{equation}
Here $\asymp$ means that the ratio of the logarithms of both sides
tends to 1 as $z\to\infty$. \eve{As discussed in \emph{Material and
    Methods}, a multiplication prefactor can be added to~(\ref{eq:4})
  but it does not affect significantly the tail of $P_T(z)$ on a
  logarithmic scale}.

In practice, the constraint $F(u(T,u_0))\ge z$ can be imposed by
adding a Lagrange multiplier term to Eq.~\ref{eq:6}, and it is easier to
use this multiplier as control parameter and simply see
\textit{a~posteriori} what value of $z$ it implies. That is to say,
perform for various values of $\lambda$ the minimization
%
\begin{equation}
  \label{eq:minimization}
  \min_{u_0} \left(\tfrac12 \|u_0\|^2_C - \lambda
  F(u(T,u_0))\right)\equiv S_T(\lambda)\,,
\end{equation}
over all $u_0$ of the form in Eq.~\ref{eq:1} (no constraint), then
observe that this implies the parametric representation
\begin{equation}
  \label{eq:2}
  I_T (z(\lambda)) = \tfrac12 \| u^\star_0(\lambda)\|^2_C, \quad
  z(\lambda) = F(u(T,u^\star_0(\lambda)))\,.
\end{equation}
where $u^\star_0(\lambda)$ denotes the minimizer obtained in
Eq.~\ref{eq:minimization}.  It is easy to see from Eqs.~\ref{eq:6}
and~\ref{eq:minimization} that $S_T(\lambda)$ is the Legendre
transform of $I_T(z)$ since:
 \begin{equation} 
   \label{eq:LD4} 
   S_T(\lambda) = \sup_{z\in
     \RR}(\lambda z - I_T(z)) = \sup_{z\in\RR}(\lambda z - \tfrac12
   \inf_{u_0\in\Omega(z)}\| u_0\|^2_C),
\end{equation}

\section{Results} We considered two sets of parameters. In Set~1 we
took $A= 5.4\cdot10^{-5}k_0^{-2}$ and $\Delta = 0.19 k_0 $. Converting
back into dimensional units using $k^{-1}_0 = 36$~m consistent with
the JONSWAP spectrum~\cite{hasselmann:1973,onorato:2001}, this implies
a significant wave height $H_s = 4\sqrt{C(0)} = 3.3$~m classified as a
\emph{rough sea}~\cite{WMO:2016}. It also yields a Benjamin-Feir index
BFI$\,=2\sqrt{2C(0)}/\Delta=0.34$,~\cite{onorato:2001,janssen:2003},
meaning that the modulational instability of a typical initial
condition is of medium intensity. In Set~2 we took
$A=3.4\cdot10^{-4}k_0^{-2}$ and $\Delta =0.19 k_0$, for which $H_s =
8.2$~m is that of a \emph{high sea} and the BFI is 0.85, meaning that
the modulational instability of a typical initial condition is
stronger.

Fig.~\ref{fig:1} (top) shows the time evolution of $|u(t,x)|$ starting
from an initial condition from Set 1 optimized so that
$\max_x |u(T,x)|= 8$~m at $T=20$~min. For comparison, Fig.~\ref{fig:1}
(bottom) shows $|u(t,x)|$ for a typical initial condition drawn from
its Gaussian distribution. To illustrate what is special about the
initial conditions identified by our optimization procedure, in
Fig.~\ref{fig:2} we show snapshots of the surface elevation
$\eta(t,x)$ at three different times, $t=0,10,20$~min (black lines),
using the constraint that $\max_x |u(T,x)|\ge 4.8$~m at
$T=20$~min. Additionally, we averaged all Monte-Carlo samples
achieving $\max_x |u(t,x)|\ge 4.8$~m, translated to the
origin. Snapshots of this mean configuration are shown in
Fig.~\ref{fig:2}~(blue lines). They agree well with those of the
optimized solution (black lines). The one standard deviation spread
around the mean Monte-Carlo realization (light blue) is reasonably
small, especially around the rogue wave at final time.  This indicates
that the event $\max_x |u(T,x)|\ge 4.8$~m is indeed realized with
probability close to 1 by starting from the most likely initial
condition consistent with this event, as predicted by LDT. The
usefulness of LDT is confirmed in Figs.~\ref{fig:3}~(a,b) depicting
the probabilities of $\max_x |u(T,x)|$ for both Sets~1 and 2
calculated via LDT optimization (lines), compared to Monte-Carlo
sampling (dots). As can be seen, the agreement is remarkable,
especially in the tail corresponding to the rogue wave regime. As
expected, the Monte-Carlo sampling becomes inaccurate in the tail,
since there the probabilities are dominated by unlikely events. The
LDT calculation, in contrast, \eve{remains efficient and accurate far
  in the tail.}

\begin{figure}
  \begin{center}
    \includegraphics[width=0.5\columnwidth]{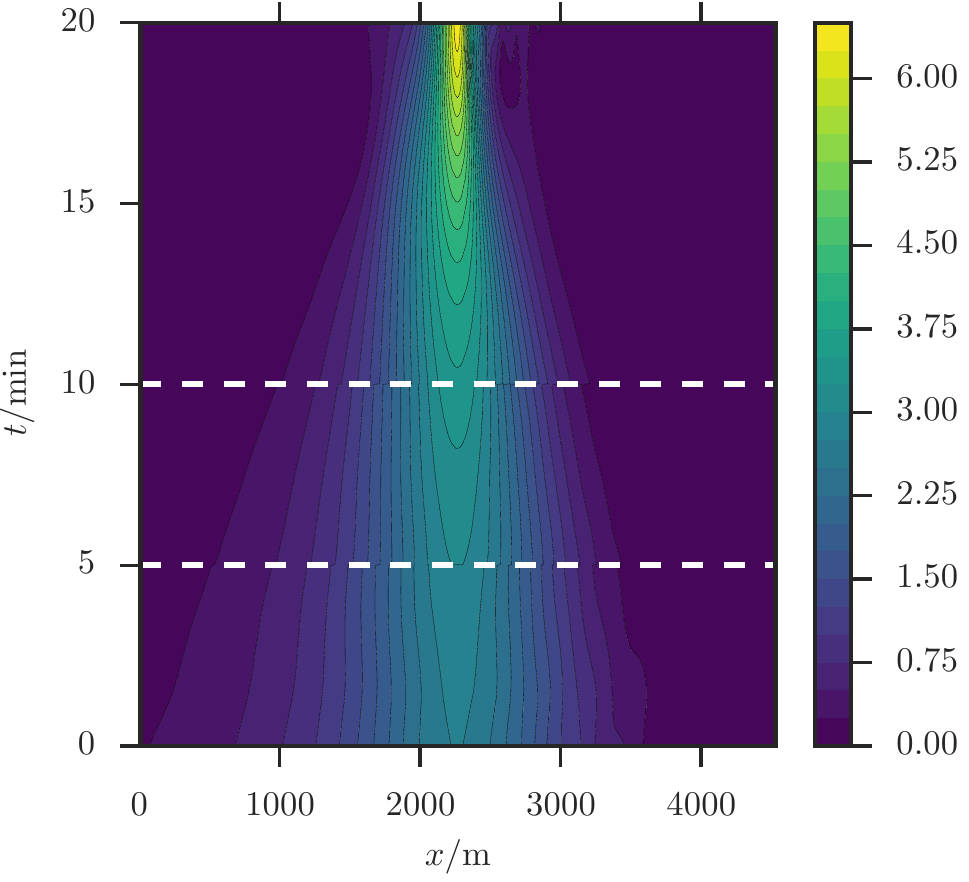}
  \end{center}
  \caption{\label{fig:4} Contourplot of the optimal trajectories from
    LDT for $T=10$, 15, and 20~min in Set~1. The trajectories,
    superposed to match at $t=T$, coincide, which is consistent with
    the convergence of the probabilities $P_T(z)$ for large $T$. }
\end{figure}

The probabilities plotted in Fig.~\ref{fig:3}~(a,b) show several
remarkable features. First, they indicate that, as $T$ gets larger,
their tails fatten significantly. For example, in Set~1
$P_{T=20\,\text{min}}(6\,\text{m})\approx 10^{-5}$, which is $5$
orders of magnitude larger than initially,
$P_{t=0\,\text{min}}(6\,\text{m})\approx 10^{-10}$. Secondly, the
probabilities converge to a limiting density for large $T$. This
occurs after some decorrelation time ~$\tau_c \approx 10$~min in Set~1
and $\tau_c \approx 3$~min in Set~2. Similarly, the LDT results
converge. In fact, this convergence can be observed at the level of
the trajectories generated from the optimal $u_0^\star$. As
Fig.~\ref{fig:4} shows, reading these trajectories backward from
$t=T$, their end portions coincide, regardless on whether $T=20$~min,
$T=15$~min, or $T=10$~min. \eve{The implications of these
  observations, in particular on the mechanism of creation of rogue
  waves and their probability of appearance within a time window, will
  be discussed in \textit{Interpretation} below.}

\eve{Before doing so, let us discuss the scalability of our results to
  larger domain sizes, referring the reader to the \textit{Supporting
    Information} for more details. As shown above, the optimization
  procedure based on large deviation theory predicts that the most
  likely way a rogue wave will occur in the domain is via the
  apparition of a single large \tg{peak} in~$|u(t,x)|$. In the set-up
  considered before, this prediction is confirmed by the brute-force
  simulations using Monte-Carlo sampling. \tg{It is clear, however,
    that for increased domain size, e.g.~by taking a domain size of
    $NL$ with $N\gg 1$, it will become increasingly likely to observe
    multiple peaks, for the simple reason that large waves can occur
    independently at multiple sufficiently separated locations.} In
  these large domains, the large deviation predictions remain valid if
  we look at the maximum of $|u(t,x)|$ in observation windows that are
  not too large (that is, about the size of the domain $L$ considered
  above). However, they deteriorate if we consider this maximum in the
  entire domain of size $NL$, in the sense that the value
  $\PP\left(\max_{x\in [0,NL]}|u(t,x)|\ge z\right)$ predicted by large
  deviation theory matches that from Monte-Carlo sampling at values of
  $z$ that are pushed further away in the tails. This is an
  \tg{entropic} effect, which is easy to correct for: events in
  different subwindows must be considered independent, and their
  probabilities superposed. That is, if we denote by
  \begin{equation}
    \label{eq:10}
    P^{N}_T(z) = \PP\Big(\max _{x\in [0,NL]} |u(T,x)| \ge z\Big),
  \end{equation}
  it can be related to
  $P_T(z) = \PP(\max _{x\in [0,L]} |u(T,x)| \ge z)$ via
\begin{equation}
  \label{eq:11}
  P^{N}_T(z) = 1 - \left(1-P_T(z)\right)^N.
\end{equation}
This formula is derived in the \textit{Supporting Information} and
shown to accurately explain the numerical results. \gd{For efficiency
  $L$ is chosen to be the smallest domain size for which boundary
  effects can be neglected, in the sense that the shape of the optimal
  trajectories does no longer change if $L$ is increased
  further.} In effect, this provides us with a method to scale up our
results to arbitrary large observation windows.}

\section{Interpretation}

The convergence of $P_T(z)$ towards a limiting function $P(z)$ has
important consequences for the significance and interpretation of our
method and its results. Notice first that this convergence can be
explained if we assume that the probability distribution of the
solutions to Eq.~\ref{eq:MNLSnondim} with Gaussian initial data
converges to an invariant measure. In this case, for large $T$, the
Monte-Carlo simulations will sample the value of $\max _x |u|$ on this
invariant measure, and the optimization procedure based on LDT will do
the same.  The timescale $\tau_c$ over which convergence occurs
depends on how far this invariant measure is from the initial Gaussian
measure of $u_0(x)$. Interestingly the values we observe for $\tau_c$
are in rough agreement with the time scales predicted by the
\gd{semi-classical limit of NLS that describes high-power pulse
  propagation}~\cite{bertola2013universality,
  tikan2017universality}. As recalled in the \textit{Supporting
  Information}, this approach predicts that the timescale of
apparition of a focusing solution starting from a large initial pulse
of maximal amplitude $U_i$ and length-scale $L_i$ is
$\tau_c = \sqrt{T_{\text{nl}}T_{\text{lin}}}$, where
$T_{\text{nl}} =\left(\tfrac12 \omega_0 k_0^2 U_i^2\right)^{-1}$ is
the nonlinear timescale for modulational instability and
$T_{\text{lin}} = 8 \omega_0^{-1} k_0^2 L_i^2$ is the linear timescale
associated to group dispersion. Setting $U_i=H_s$ (the size at the
onset of rogue waves) and $L_i=\sqrt{2\pi}\Delta^{-1}$ (the
correlation length of the initial field) gives $\tau_c\simeq 18$~min
for Set~1 and $\tau_c\simeq 8$~min for Set~2, consistent with the
convergence times of $P_T(z)$. This observation has implications in
terms of the mechanism of apparition of rogue waves, in particular
their connection to the so-called Peregrine soliton, that has been
invoked as prototype mechanism for rogue waves
creation~\cite{peregrine:1983, akhmediev2009waves, shrira2010makes,
  akhmediev2013recent, onorato-residori-bortolozzo-etal:2013,
  toenger2015emergent}, in particular for water waves
\cite{chabchoub2011rogue, chabchoub2012super, chabchoub2016tracking},
plasmas \cite{bailung2011observation} and fiber optics
\cite{kibler2010peregrine, suret2016single,
  tikan2017universality}. This connection is discussed in the
\textit{Supporting Information}.

\begin{figure}
  \begin{center}
    \includegraphics[width=0.6\columnwidth]{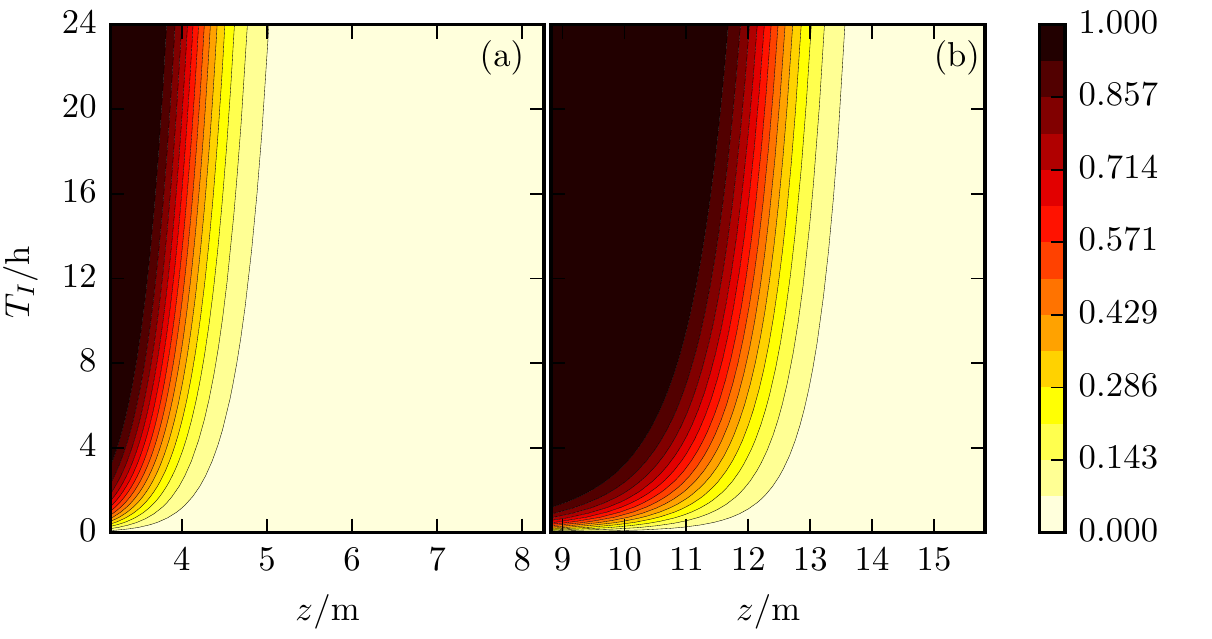}
  \end{center}
  \caption{\label{fig:5} Contourplots of the probability to observe a
    wave whose amplitude exceeds~$z$ in the time window $[0,T_I]$ for
    Sets~1 (a) and 2~(b). }
\end{figure}

Our findings also indicate that, even though the assumption that
$u_0(x)$ is Gaussian is incorrect in the tail (that is, $P_{T=0}(z)$
is not equal to the limiting $P(z)$ in the tail), it contains the
right seeds to estimate $P(z)$ via $P_T(z)$ if
$T\gtrsim \tau_c$\footnote{This convergence occurs on the timescale
  $\tau_c$ which is much smaller than the mixing time for the
  solutions of Eq.~\ref{eq:MNLSnondim}, i.e.~ the time it would take
  from a given initial condition, rather than an ensemble thereof, to
  sample the invariant measure.}.  Altogether this is consistent with
the scenario put forward by Sapsis and collaborators
in~\cite{mohamad2016probabilistic, farazmand-sapsis:2017} to explain
how extreme events arise in intermittent dynamical systems and
calculate their probability: they occur when the system hits small
instability pockets which trigger a large transient excursion. In this
scenario, as long as the initial probability distribution in these
pockets is accurate, the dynamics will permit precise estimation of
the distribution tail. In some sense, the distribution of the initial
condition plays a role of the prior distribution in Bayesian
inference\footnote{Note in particular that the Gaussian field in
  Eq.~\ref{eq:1} is the random field that maximizes entropy given the
  constraint on its covariance $C(x)$.}, and the posterior can be
effectively sampled by adding the additional information from the
dynamics over short periods of time during which instabilities can
occur. In~\cite{mohamad2016probabilistic}, this picture was made
predictive by using a two-dimensional ansatz for the initial
condition~$u_0(x)$ \eve{to avoid having to perform sampling in
  high-dimension over the original $u_0(x)$}. What our results show is
that this approximation can be avoided altogether by using LDT to
perform the calculations directly with the full Gaussian initial
condition in Eq.~\ref{eq:1}.

Interestingly, we can use the results above to calculate the
probability of occurrence of rogue waves in a given time window. More
precisely, the probability $p(z,T_I) $ that \eve{a rogue wave of amplitude
larger than $z$ be observed in the domain $[0,L]$ during $[0,T_I]$
(i.e.~that $\max_{t\in [0,T_I]}\max_{x\in [0,L]} |u(t,x)| \ge z $)
can be estimated in terms of $P(z)$ and $\tau_c$ as
\begin{equation}
  \label{eq:8}
  p  \equiv \PP\Big( \max_{t\in [0,T_I]}\max_{x\in
    [0,L]} |u(t,x)| \ge z\Big) \sim 1- (1-P(z) )^{T_I/\tau_c}\,,
\end{equation}
}where we used the fact that rogue waves can be considered independent
on timescales larger than $\tau_c$ and assumed $T_I\gg\tau_c$.  The
function $p$ is plotted in Fig.~\ref{fig:5} as a function of $z$ and
$T_I$.  For example for Set~1, Eq.~\ref{eq:8} indicates a 50\% chance
to observe a rogue wave of height $z=4$~m (that is, about 8~m from
crest-to-trough) after 11 hours (using $\tau_c= 10$~min and
$P(z=4~\text{m}) = 1.1\cdot10^{-2}$); if we wait 30 hours, the chance
goes up to 85\%. Similarly, for Set~2 the chance to observe a wave of
11~m height is about 50\% after 3~hours and about 85\% after 8~hours
($\tau_c= 3$~min and $P(z=11~\text{m}) = 1.2\cdot10^{-2}$).

\section{Concluding remarks}

We have shown how an optimization problem building on LDT can be used
to predict the pathway and likelihood of appearance of rogue waves in
the solutions of MNLS fed by random initial data consistent with
observations. This setup guarantees accuracy of the core of the
initial distribution, which in turn permits the precise estimation of
its tail via the dynamics. Our results give quantitative estimate for
the probabilities of observing high amplitude waves within a given
time window. These results also show that rogue waves have very
specific precursors, \eve{a feature that was already noted in
  \cite{farazmand2017reduced} in the context of a reduced model and
  could potentially be used for their early detection.}

\section{Materials and Methods}

\subsection{Laplace method and large deviations}

For the reader's convenience, here we recall some standard large
deviations results that rely on the evaluation of Gaussian integrals by
Laplace's method and are at the core of the method we propose. It is
convenient to rephrase the problem abstractly and consider the
estimation of
\begin{equation}
  \label{eq:7}
  P(z) = \PP(\phi(\theta)> z)\,,
\end{equation}
where $\theta\in \RR^D$ are Gaussian random variables with mean zero
and covariance $\text{Id}$, and $\phi: \RR^D \to \RR$ is some real
valued function -- as long as we truncate the sum in Eq.~\ref{eq:1}
to a finite number of modes, $|n|\le M$, the problem treated in this
paper can be cast in this way, with $\theta$ playing the role of
$C^{-1/2}u_0$ and $\phi(\cdot)$ that of $F(u(T,C^{-1/2}\cdot))$. The
probability $P(z)$ in Eq.~\ref{eq:7} is given by
\begin{equation}
  \label{eq:5}
  P(z) = (2\pi)^{-D/2} \int_{\Omega(z)}
  e^{-\frac12 |\theta|^2}d\theta\,,
\end{equation}
where $\Omega(z)= \{\theta : \phi(\theta) >z\}$. The interesting
case is when this set does not contain the origin,
$0 \not \in \Omega(z)$, which we will assume is true when
$z>0$. We also make two additional assumptions:
\begin{enumerate}
\item The point on the boundary $\partial \Omega(z)$ that is
  closest to the origin is isolated:   Denoting this point as
\begin{equation}
  \label{eq:mintheta}
  \theta^\star(z) = \argmin_{\theta\in \partial \Omega(z)} |\theta|^2\,,
\end{equation}
we assume that
\begin{equation}
  \label{eq:assumptions}
  \begin{aligned}[c]
    &\text{$\tfrac12|\theta^\star(z)|^2$ is strictly
      increasing with $z\ge0$\,; } \\
    &\lim_{z\to\infty} \tfrac12|\theta^\star(z)|^2= \infty\,.
  \end{aligned}
\end{equation}
\item The connected piece of~$\partial \Omega(z)$ that
  contains~$\theta^\star(z)$ is smooth with a curvature that is
  bounded by a constant independent of~$z$.
\end{enumerate}
The point $\theta^\star(z)$ satisfies the Euler-Lagrange equation for
Eq.~\ref{eq:mintheta}, with the constraint incorporated via a Lagrange
multiplier term:
\begin{equation}
  \label{eq:ELc}
  \theta^\star(z) = \lambda \nabla \phi (\theta^\star(z))
\end{equation}
for some Lagrange multiplier $\lambda$. This implies that
\begin{equation}
  \label{eq:parallel}
  \frac{\theta^\star(z)}{|\theta^\star(z)|}= 
  \frac{\nabla \phi (\theta^\star(z))}{|\nabla \phi (\theta^\star(z))|} =
  \hat n(z)\,.
\end{equation}
where $\hat n(z)$ denotes the inward pointing unit vector normal to
$\partial \Omega(z)$ at $\theta^\star(z)$.  If we move inside the
set $\Omega(z)$ from $\theta^\star(z)$ in the direction of
$\hat n(z)$, the norm $|\theta|^2 $ increases under the assumptions in
Eq.~\ref{eq:assumptions}. Indeed, setting
$\theta = \theta^\star(z)+\hat n(z) u$ with $u\ge0$, we have
\begin{equation}
  \label{eq:away}
  \begin{aligned}
    |\theta|^2 &= |\theta^\star(z) |^2 + 2 \langle\hat n(z)
    ,\theta^\star(z) \rangle u + u^2\\
    & = |\theta^\star(z) |^2 + 2 |\theta^\star(z) | z + z^2\,,
  \end{aligned}
\end{equation}
where we used Eq.~\ref{eq:parallel}. In fact, if we were to perform
the integral in that direction, the natural variable of integration
would be to rescale $u \to u/|\theta^\star(z) |$. In particular, if we
were to replace $\Omega(z)$ by the half space
$P(z) = \{\theta \, | \, \hat n(z)\cdot(\theta-\theta^*(z))>0\}$, it
would be easy to estimate the integral in Eq.~\ref{eq:5} by
introducing a local coordinate system around $\theta^*(z)$, whose
first coordinate is in the direction of $\hat n(z)$. Indeed this would
give:
\begin{align}
  & (2\pi) ^{-D/2}\int_{P(z)} e^{-\tfrac12|\theta|^2} d\theta\nonumber\\
  & = (2\pi) ^{-D/2} \int_0^\infty e^{-\tfrac12
    |\theta^\star(z)|^2- |\theta^\star(z) | u - \tfrac12u^2} du
    \ \int_{\RR^{N-1}} \!\!\!\!e^{-\tfrac12
    |\eta|^2} d\eta\nonumber\\
  & = (2\pi) ^{-1/2}e^{-\tfrac12 |\theta^\star(z)|^2} \int_0^\infty
    e^{- |\theta^\star(z) | u - \tfrac12u^2} du\nonumber\\ 
  & = (2\pi) ^{-1/2}
    |\theta^\star(z) | ^{-1} e^{-\tfrac12 |\theta^\star(z)|^2}
    \int_0^\infty e^{- v - \tfrac12 |\theta^\star(z) | ^{-2} v^2} dv
  \nonumber\\ 
  & \sim (2\pi) ^{-1/2} |\theta^\star(z) | ^{-1} e^{-\tfrac12
    |\theta^\star(z)|^2}\qquad\qquad \text{as\ $z \to \infty$.}
    \label{eq:planeest}
\end{align}
The last approximation goes beyond a large deviations estimate
\eve{(i.e. it includes the prefactor)}, and it
implies
\begin{equation}
  \label{eq:ldplane2}
  \lim_{z\to\infty} |\theta^\star(z) | ^{-2}  \log\left( (2\pi) ^{-D/2}\int_{P(z)}
  e^{-\tfrac12|\theta|^2} d\theta\right)  = -\frac12\,.
\end{equation}
This log-asymptotic estimate is often written as
\begin{equation}
  \label{eq:ldplane}
    \int_{P(z)} e^{-\tfrac12|\theta|^2} d\theta\\
    \asymp e^{-\tfrac12
  |\theta^\star(z)|^2}\qquad\text{as\ $z \to \infty$.}
\end{equation}
Interestingly, while the asymptotic estimate in Eq.~\ref{eq:planeest}
does not necessarily apply to the original integral in Eq.~\ref{eq:5}
\eve{(that is, the prefactor may take different forms depending on the
  shape of $\partial \Omega(z)$ near $\theta^\star(z)$)}, the rougher
log-asymptotic estimate in Eq.~\ref{eq:ldplane} does as long as the
the boundary~$\partial \Omega(z)$ is smooth, with a curvature that is
bounded by a constant independent of~$z$. This is because because the
contribution (positive or negative) to the integral over the region
between $\Omega(z)$ and $P(z)$ is subdominant in that case, in the
sense that the log of its amplitude is dominated by
$|\theta^\star(z)|$. This is the essence of the large deviations
result that we apply in this paper.

\subsection{Numerical aspects} 

To perform the calculations, we solved Eq.~\ref{eq:MNLSnondim} with
$L=40\pi$ \eve{and periodic boundary conditions, and checked that this
  domain is large enough to make the effect of these boundary
  conditions negligible (see \textit{Supporting Information}).  The
  spatial domain was discretized using $2^{12}$ equidistant
  gridpoints, which is enough to resolve the solution of
  Eq.~\ref{eq:MNLSnondim}. To evolve the field $u(t,x)$ in time we
  used a pseudo-spectral second order exponential time-differencing
  (ETD2RK) method~\cite{cox:2002, kassam:2005}.

\eve{When performing the Monte-Carlo simulations, we used $10^6$
  realizations of the random initial data constructed by truncating
  the sum in Eq.~\ref{eq:1} over the $M=23$ modes with
  $-11\le n\le11$, i.e $-3\Delta \le k_n \le 3 \Delta$: these modes
  carry most of the variance, and we checked that adding more modes to
  the initial condition did not affect the results in any significant
  way (see \textit{Supporting
    Information}). 
}


\subsection{Optimization procedure}

\eve{As explained above, the large deviation rate function $I_T(z)$ in
  Eq.~\ref{eq:6} can be evaluated by solving the dual optimization
  problem in Eq.~\ref{eq:minimization}, which we rewrite as
  $S_T(\lambda) = \inf_{u_0} E(u_0,\lambda)$, where we defined the
  cost function
\begin{equation} \label{eq:met1}
    E(u_0,\lambda) \equiv \tfrac12 \|u_0\|_C^2 - \lambda F(u(T,u_0))\,.
\end{equation}
We performed this minimization using steepest descent with adaptive
step (line-search) and preconditioning of the
gradient~\cite{borzi2011computational}. This involves evaluating the
(functional) gradient of $E_T(u_0,\lambda)$ with respect to $u_0$.
Using the chain rule, this gradient can be expressed as (using
compact vectorial notation)
\begin{equation}
  \label{eq:met3}
  \frac{\delta E}{\delta u_0} = C^{-1} u_0 - \lambda  J^T(T,u_0) 
  \frac{\delta F}{\delta u}
\end{equation}
where $J(t,u_0) = \delta u (t,u_0)/\delta u_0$ is the Jacobian of the
transformation $u_0\to u(t,u_0)$. Collecting all terms on the
right-hand-side of the MNLS Eq.~\ref{eq:MNLSnondim} into $b(u)$, this
equation can be written as
\begin{equation}
  \label{eq:met4}
 \partial_{t} u = b(u),  \quad u(t=0) = u_0\,,
\end{equation}
and it is easy to see that in this notation $J(t,u_0)$ satisfies
\begin{equation}
  \label{eq:met5}
  \partial_t J = \frac{\delta b}{\delta u} J, \qquad J(t=0) = \text{Id}.
\end{equation}

Consistent with what was done in the Monte-Carlo sampling, to get the
results presented above we truncated the initial data $u_0$ over
$M=23$ modes using the representation
\begin{equation}
  \label{eq:9}
  u_0(x) = \sum_{n=-11}^{11} e^{ik_n x} \hat a_n, \quad k_n = 2\pi n/L.
\end{equation}
This means that minimization of Eq.~\ref{eq:met1} was performed in the
$2M-1=45$ dimensional space spanned by the modes $\hat a_n$,
accounting for invariance by an overall phase shift -- to check
convergence we also repeated this calculation using larger values of
$M$ and found no noticeable difference in the results (see
\textit{Supporting Information}). 

In practice, the evaluation of the gradient in Eq.~\ref{eq:met3} was
performed by integrating both $u(t)$ and $J(t)$ up to time $t=T$.
Eq.~\ref{eq:met5} was integrated using the same pseudo-spectral method
as for Eq.~\ref{eq:MNLSnondim} on the same grid. To perform the
steepest descent step, we then preconditioned the gradient through
scalar multiplication by the step-independent, diagonal metric with
the components of the spectrum $\hat C_n$ as diagonal elements. 

}

}

\subsubsection*{Acknowledgment}

We thank W.~Craig and M.~Onorato for helpful discussions, and
O.~B\"uhler, M.~Mohamad, and T.~Sapsis for interesting
comments. \gd{We also thank the anonymous reviewer for drawing our
  attention to the semi-classical theory for the Nonlinear
  Schr\"odinger Equation.} GD is supported by the joint Math PhD
program of Politecnico and Universit\`a di Torino.  EVE is supported
in part by the Materials Research Science and Engineering Center
(MRSEC) program of the National Science Foundation (NSF) under award
number DMR-1420073 and by NSF under award number DMS-1522767.



\begin{thebibliography}{10}

\bibitem{muller2005rogue}
M{\"u}ller P, Garrett C, Osborne A (2005) Rogue waves.
\newblock {\em Oceanography} 18(3):66.

\bibitem{white1998chance}
White BS, Fornberg B (1998) On the chance of freak waves at sea.
\newblock {\em J. Fluid Mech.} 355:113--138.

\bibitem{haver2004possible}
Haver S (2004) A possible freak wave event measured at the draupner jacket
  january 1 1995.
\newblock {\em Rogue waves 2004} pp. 1--8.

\bibitem{nikolkina2011rogue}
Nikolkina I, Didenkulova I (2011) Rogue waves in 2006--2010.
\newblock {\em Nat. Hazards Earth Syst. Sci.} 11(11):2913--2924.

\bibitem{onorato-residori-bortolozzo-etal:2013}
Onorato M, Residori S, Bortolozzo U, Montina A, Arecchi F (2013) Rogue waves
  and their generating mechanisms in different physical contexts.
\newblock {\em Phys. Rep.} 528(2):47--89.

\bibitem{nazarenko:2016}
Nazarenko S, Lukaschuk S (2016) Wave turbulence on water surface.
\newblock {\em Annu. Rev. Condens. Matter Phys.} 7:61--88.

\bibitem{akhmediev2009extreme}
Akhmediev N, Soto-Crespo JM, Ankiewicz A (2009) Extreme waves that appear from
  nowhere: on the nature of rogue waves.
\newblock {\em Physics Letters A} 373(25):2137--2145.

\bibitem{akhmediev2010editorial}
Akhmediev N, Pelinovsky E (2010) Editorial--introductory remarks on
``discussion \& debate: Rogue waves--towards a unifying concept?''.
\newblock {\em Eur. Phys. J. Special Topics} 185(1):1--4.

\bibitem{onorato2016origin}
Onorato M, Proment D, El G, Randoux S, Suret P (2016) On the origin of
  heavy-tail statistics in equations of the nonlinear schr{\"o}dinger type.
\newblock {\em Phys. Lett. A} 380(39):3173--3177.

\bibitem{benjamin-feir:1967}
Benjamin TB, Feir JE (1967) The disintegration of wave trains on deep water
  {Part} 1. {Theory}.
\newblock {\em J. Fluid Mech.} 27(03):417--430.

\bibitem{zakharov:1968}
Zakharov VE (1968) Stability of periodic waves of finite amplitude on the
  surface of a deep fluid.
\newblock {\em J. Appl. Mech. Tech. Phys.} 9(2):190--194.

\bibitem{kuznetsov:1977}
Kuznetsov EA (1977) Solitons in a parametrically unstable plasma.
\newblock {\em Akademiia Nauk SSSR Doklady} 236:575--577.

\bibitem{peregrine:1983}
Peregrine DH (1983) Water waves, nonlinear {Schr\"odinger} equations and their
  solutions.
\newblock {\em The ANZIAM Journal} 25(01):16--43.

\bibitem{akhmediev1987exact}
Akhmediev N, Eleonskii V, Kulagin N (1987) Exact first-order solutions of the
  nonlinear schr{\"o}dinger equation.
\newblock {\em Theoretical and mathematical physics} 72(2):809--818.

\bibitem{osborne2000nonlinear}
Osborne AR, Onorato M, Serio M (2000) The nonlinear dynamics of rogue waves and
  holes in deep-water gravity wave trains.
\newblock {\em Phys. Lett. A} 275(5):386--393.

\bibitem{zakharov-ostrovsky:2009}
Zakharov VE, Ostrovsky LA (2009) Modulation instability: {The} beginning.
\newblock {\em Physica D} 238(5):540--548.

\bibitem{onorato:2009}
Onorato M, et~al. (2009) Statistical properties of directional ocean waves: the
  role of the modulational instability in the formation of extreme events.
\newblock {\em Phys. Rev. Lett.} 102(11):114502.

\bibitem{dysthe:1979}
Dysthe KB (1979) Note on a {Modification} to the {Nonlinear} {Schr\"odinger}
  {Equation} for {Application} to {Deep} {Water} {Waves}.
\newblock {\em Proc. R. Soc. Lond. A} 369(1736):105--114.

\bibitem{stiassnie:1984}
Stiassnie M (1984) Note on the modified nonlinear schr{\"o}dinger equation for
  deep water waves.
\newblock {\em Wave motion} 6(4):431--433.

\bibitem{trulsen1996modified}
Trulsen K, Dysthe KB (1996) A modified nonlinear schr{\"o}dinger equation for
  broader bandwidth gravity waves on deep water.
\newblock {\em Wave motion} 24(3):281--289.

\bibitem{craig2010hamiltonian}
Craig W, Guyenne P, Sulem C (2010) A hamiltonian approach to nonlinear
  modulation of surface water waves.
\newblock {\em Wave Motion} 47(8):552--563.

\bibitem{gramstad2011hamiltonian}
Gramstad O, Trulsen K (2011) Hamiltonian form of the modified nonlinear
  schr{\"o}dinger equation for gravity waves on arbitrary depth.
\newblock {\em J. Fluid Mech.} 670:404--426.

\bibitem{onorato2004observation}
Onorato M, et~al. (2004) Observation of strongly non-gaussian statistics for
  random sea surface gravity waves in wave flume experiments.
\newblock {\em Phys. Rev. E} 70(6):067302.

\bibitem{chabchoub2011rogue}
Chabchoub A, Hoffmann N, Akhmediev N (2011) Rogue wave observation in a water
  wave tank.
\newblock {\em Phys. Rev. Lett.} 106(20):204502.

\bibitem{chabchoub2012super}
Chabchoub A, Hoffmann N, Onorato M, Akhmediev N (2012) Super rogue waves:
  observation of a higher-order breather in water waves.
\newblock {\em Phys. Rev. X} 2(1):011015.

\bibitem{goullet2011numerical}
Goullet A, Choi W (2011) A numerical and experimental study on the nonlinear
  evolution of long-crested irregular waves.
\newblock {\em Phys. Fluids} 23(1):016601.

\bibitem{lo1985numerical}
Lo E, Mei CC (1985) A numerical study of water-wave modulation based on a
  higher-order nonlinear schr{\"o}dinger equation.
\newblock {\em J. Fluid Mech.} 150:395--416.

\bibitem{cousins-sapsis:2015A}
Cousins W, Sapsis TP (2015) Unsteady evolution of localized unidirectional
  deep-water wave groups.
\newblock {\em Phys. Rev. E} 91(6):063204.

\bibitem{cousins2016reduced}
Cousins W, Sapsis TP (2016) Reduced-order precursors of rare events in
  unidirectional nonlinear water waves.
\newblock {\em J. Fluid Mech.} 790:368--388.

\bibitem{nazarenko2011wave}
Nazarenko S (2011) {\em Wave turbulence}.
\newblock (Springer Science \& Business Media) Vol.{} 825.

\bibitem{hasselmann:1973}
Hasselmann K, et~al. (1973) Measurements of wind-wave growth and swell decay
  during the joint north sea wave project (jonswap), (Deutches Hydrographisches
  Institut), Technical report.

\bibitem{onorato:2001}
Onorato M, Osborne AR, Serio M, Bertone S (2001) Freak waves in random oceanic
  sea states.
\newblock {\em Phys. Rev. Lett.} 86(25):5831.

\bibitem{WMO:2016}
Organization WM, ed. (2016) {\em Manual on {Codes} - {International} {Codes},
  {Volume} {I}.1, {Annex} {II} to the {WMO} {Technical} {Regulations}: part
  {A}- {Alphanumeric} {Codes}}, {WMO}- {No}. 306.
\newblock (Secretariat of the World Meteorological Organization).

\bibitem{janssen:2003}
Janssen PAEM (2003) Nonlinear {Four}-{Wave} {Interactions} and {Freak} {Waves}.
\newblock {\em J. Phys. Oceanogr.} 33(4):863--884.

\bibitem{bertola2013universality}
Bertola M, Tovbis A (2013) Universality for the focusing nonlinear
  schr{\"o}dinger equation at the gradient catastrophe point: rational
  breathers and poles of the tritronqu{\'e}e solution to painlev{\'e} i.
\newblock {\em Communications on Pure and Applied Mathematics} 66(5):678--752.

\bibitem{tikan2017universality}
Tikan A, et~al. (2017) Universality of the peregrine soliton in the focusing
  dynamics of the cubic nonlinear schr{\"o}dinger equation.
\newblock {\em Physical Review Letters} 119(3):033901.

\bibitem{akhmediev2009waves}
Akhmediev N, Ankiewicz A, Taki M (2009) Waves that appear from nowhere and
  disappear without a trace.
\newblock {\em Physics Letters A} 373(6):675--678.

\bibitem{shrira2010makes}
Shrira VI, Geogjaev VV (2010) What makes the peregrine soliton so special as a
  prototype of freak waves?
\newblock {\em Journal of Engineering Mathematics} 67(1):11--22.

\bibitem{akhmediev2013recent}
Akhmediev N, Dudley JM, Solli D, Turitsyn S (2013) Recent progress in
  investigating optical rogue waves.
\newblock {\em Journal of Optics} 15(6):060201.

\bibitem{toenger2015emergent}
Toenger S, et~al. (2015) Emergent rogue wave structures and statistics in
  spontaneous modulation instability.
\newblock {\em Scientific reports} 5.

\bibitem{chabchoub2016tracking}
Chabchoub A (2016) Tracking breather dynamics in irregular sea state
  conditions.
\newblock {\em Physical review letters} 117(14):144103.

\bibitem{bailung2011observation}
Bailung H, Sharma S, Nakamura Y (2011) Observation of peregrine solitons in a
  multicomponent plasma with negative ions.
\newblock {\em Physical review letters} 107(25):255005.

\bibitem{kibler2010peregrine}
Kibler B, et~al. (2010) The peregrine soliton in nonlinear fibre optics.
\newblock {\em Nature Physics} 6(10):790.

\bibitem{suret2016single}
Suret P, et~al. (2016) Single-shot observation of optical rogue waves in
  integrable turbulence using time microscopy.
\newblock {\em Nature communications} 7.

\bibitem{mohamad2016probabilistic}
Mohamad MA, Cousins W, Sapsis TP (2016) A probabilistic decomposition-synthesis
  method for the quantification of rare events due to internal instabilities.
\newblock {\em J. Comp. Phys.} 322:288--308.

\bibitem{farazmand-sapsis:2017}
Farazmand M, Sapsis TP (2017) A variational approach to probing extreme events
  in turbulent dynamical systems.
\newblock {\em arXiv preprint arXiv:1704.04116}.

\bibitem{farazmand2017reduced}
Farazmand M, Sapsis TP (2017) Reduced-order prediction of rogue waves in
  two-dimensional deep-water waves.
\newblock {\em Journal of Computational Physics} 340:418--434.


\bibitem{cox:2002}
Cox SM, Matthews PC (2002) Exponential time differencing for stiff systems.
\newblock {\em J. Comp. Phys.} 176(2):430--455.

\bibitem{kassam:2005}
Kassam AK, Trefethen LN (2005) Fourth-order time-stepping for stiff pdes.
\newblock {\em SIAM J. Sci. Comput.} 26(4):1214--1233.

\bibitem{borzi2011computational}
Borz{\`\i} A, Schulz V (2011) {\em Computational optimization of systems
  governed by partial differential equations}.
\newblock (SIAM).


\end{thebibliography}

\clearpage
\includepdf[pages={1}]{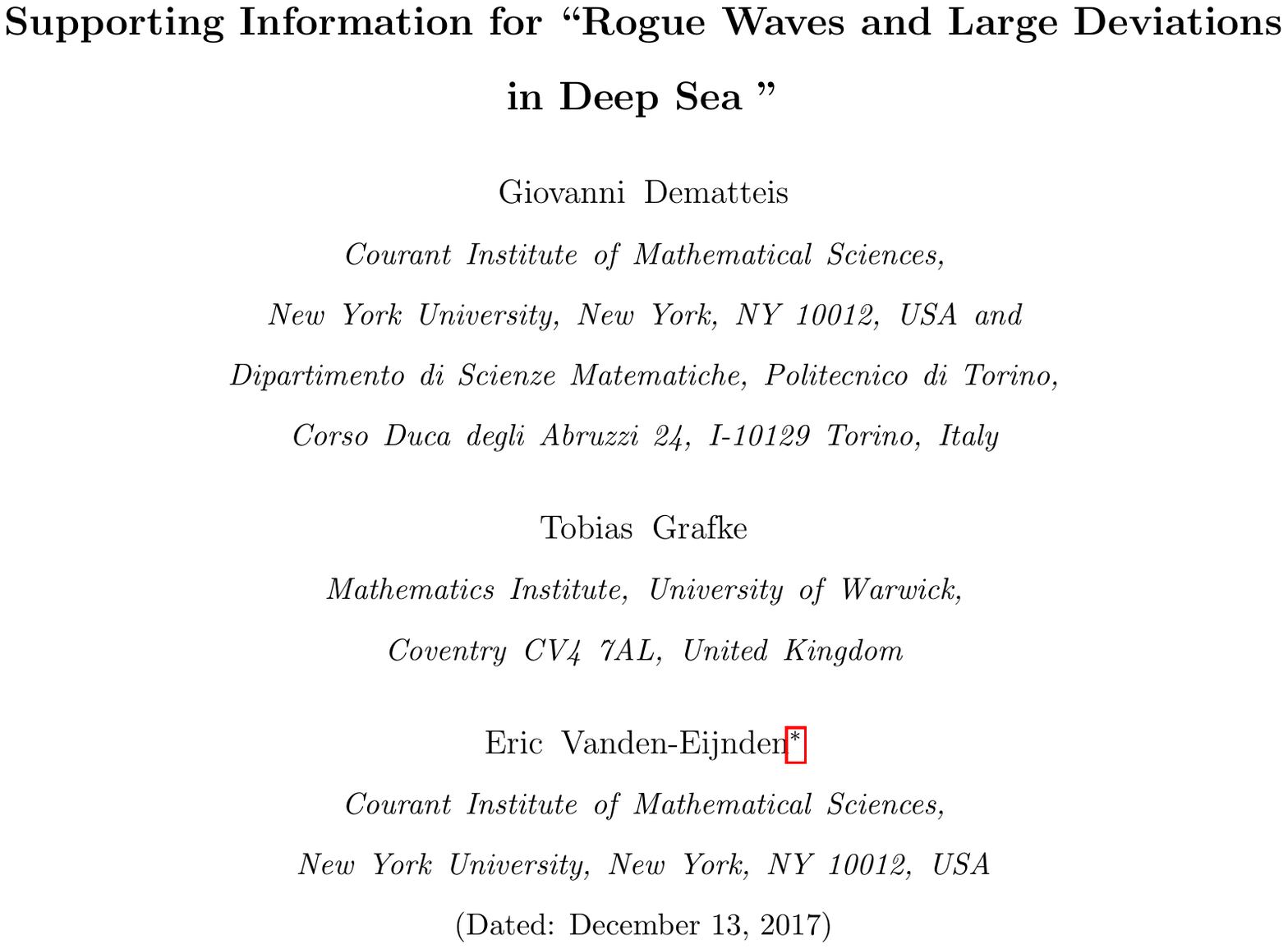}
\clearpage
\includepdf[pages={2}]{roguesuppinf.pdf}
\clearpage
\includepdf[pages={3}]{roguesuppinf.pdf}
\clearpage
\includepdf[pages={4}]{roguesuppinf.pdf}
\clearpage
\includepdf[pages={5}]{roguesuppinf.pdf}
\clearpage
\includepdf[pages={6}]{roguesuppinf.pdf}
\clearpage
\includepdf[pages={7}]{roguesuppinf.pdf}
\clearpage
\includepdf[pages={8}]{roguesuppinf.pdf}
\clearpage
\includepdf[pages={9}]{roguesuppinf.pdf}
\clearpage
\includepdf[pages={10}]{roguesuppinf.pdf}
\clearpage
\includepdf[pages={11}]{roguesuppinf.pdf}
\clearpage
\includepdf[pages={12}]{roguesuppinf.pdf}
\clearpage
\includepdf[pages={13}]{roguesuppinf.pdf}
\clearpage
\includepdf[pages={14}]{roguesuppinf.pdf}
\clearpage
\includepdf[pages={15}]{roguesuppinf.pdf}
\clearpage
\includepdf[pages={16}]{roguesuppinf.pdf}
\clearpage
\includepdf[pages={17}]{roguesuppinf.pdf}

\end{document}